\documentclass[aps, prl, twocolumn, notitlepage, superscriptaddress]{revtex4-2}
\usepackage[T1]{fontenc}
\usepackage{amsmath}
\usepackage{booktabs}
\usepackage{appendix}
\usepackage[inline]{enumitem}
\usepackage{multirow}
\usepackage{amssymb}
\usepackage{amsthm}
\usepackage{array}
\usepackage{graphicx}
\usepackage{verbatim}
\usepackage{bbm}
\usepackage{color}
\usepackage[normalem]{ulem}
\usepackage{comment}
\usepackage[colorlinks=true,citecolor=red,urlcolor=blue]{hyperref}
\DeclareGraphicsExtensions{.png,.pdf,.eps}

\def \diracspacing {0.7pt}
\newcommand{\bra}[1]{\langle #1 \hspace{\diracspacing} |} 
\newcommand{\ket}[1]{| \hspace{\diracspacing} #1 \rangle} 
\newcommand{\ketbra}[2]{| \hspace{\diracspacing} #1 \rangle \langle #2 \hspace{\diracspacing} |} 
\newcommand{\ketbraq}[1]{\ketbra{#1}{#1}} 
\newcommand{\prlparagraph}[1]{\emph{#1.---}}

\newcommand{\I}{\mathbb{I}}

\newcommand{\cH}{\mathcal{H}}
\newcommand{\cA}{\mathcal{A}}
\newcommand{\cB}{\mathcal{B}}

\newcommand{\bC}{\mathbb{C}}

\newcommand{\rA}{\mathrm{A}}
\newcommand{\rB}{\mathrm{B}}

\newcommand{\rr}{\mathbf{r}}
\newcommand{\aux}{\text{aux}}

\DeclareMathOperator{\tr}{tr}

\theoremstyle{definition}
\theoremstyle{plain}

\theoremstyle{remark}

\begin{document}
\title{Maximal global device-independent randomness from projective measurements in every dimension}
\author{M\'at\'e Farkas}
\email{mate.farkas@york.ac.uk}
\affiliation{Department of Mathematics, University of York, York, YO10 5GH, United Kingdom}

\author{Piotr Mironowicz}
\email{piotr.mironowicz@gmail.com }
\affiliation{Center for Theoretical Physics, Polish Academy of Sciences, al. Lotnik\'{o}w 32/46, 02-668 Warsaw, Poland}
\affiliation{Department of Algorithms and System Modeling, Faculty of Electronics, Telecommunications and Informatics, Gda\'{n}sk University of Technology, Poland}

\author{Remigiusz Augusiak}
\email{augusiak@cft.edu.pl}
\affiliation{Center for Quantum-Enabled Computing, Center for Theoretical Physics, Polish Academy of Sciences, al. Lotnik\'{o}w 32/46, 02-668 Warsaw, Poland}

\begin{abstract}
Device-independent random number generation (DIQRNG) is the most secure form of generating private randomness using quantum physical processes. Its strength lies in producing numbers that are impossible to predict by any eavesdropper restricted by the laws of quantum theory. Moreover, security is proven solely from observed measurement statistics, without the need to characterise or trust the devices used in random number generation. Implementing DIQRNG is, however, costly, as it requires high-quality entangled systems. It is therefore important to make the best use of available resources. In this work, we show that using projective measurements---which are most readily implementable experimentally---one can certify $2\log(d)$ bits of device-independent randomness from a bipartite system of local dimension $d$ for every $d \ge 2$, thus reaching the theoretically maximum possible rate of DIQRNG. We provide explicit protocols reaching $2\log(d)$ bits based on mutually unbiased bases. Furthermore, we compute numerical bounds on the rate for the case of imperfect implementations, showing that our protocols are robust to experimental noise.
\end{abstract}

\maketitle

\prlparagraph{Introduction}
 Device-independent random number generation (DIQRNG) makes it possible to certify the randomness of the outcomes of certain quantum measurements even if the underlying quantum state and the measurements are completely uncharacterised. As long as the standard assumptions of Bell non-locality~\cite{brunner2014bell} are satisfied, numbers generated by a DIQRNG protocol cannot be predicted by any eavesdropper limited by the laws of quantum mechanics~\cite{mayers1998quantum,Colbeck2007}. Such secure random numbers find applications in cryptography~\cite{kocc2009cryptographic}, as the security of most cryptographic protocols depends on private random numbers~\cite{pironio2010random}.

DIQRNG protocols operate in a \emph{Bell scenario}: two trusted parties share a pair of entangled quantum systems, each measure one of the systems and record their measurement choices (`settings') and outcomes~\cite{PhysicsPhysiqueFizika.1.195,PhysRevLett.23.880}. They collect the measurement statistics (often referred to as the \emph{correlation}), and infer their conclusions---in particular the private randomness of their measurement outcomes---from the observed correlation.

DIQRNG is technologically challenging~\cite{liu2018device}. Among other difficulties, scaling up the number of degrees of freedom (the \emph{dimension}) of the quantum system is complicated in practice. Since the dimension limits the amount of certifiable randomness per measurement (which is easily seen for projective measurements, see the next section), it is important to devise protocols that certify as much randomness as possible in a given dimension.

It has been shown \cite{farkas2024maximal} that the maximal possible, $2 \log(d)$ bits, of device-independent randomness can be certified by a single party (i.e.~\emph{locally}) using generalised, non-projective measurements. That work leaves several questions open, one of them being the practical question of how much device-independent randomness can be certified by \emph{both} parties (i.e.~\emph{globally})? This work answers this question for the case of projective measurements, which are the most practical to implement experimentally. In particular, we show that the maximum theoretically possible amount of global randomness using projective measurements---also $2 \log(d)$ bits---can be certified in every dimension~$d$. Up to our knowledge, so far this has only been known for the case of $d=2$ \cite{AMP12,WKBS+20}. Furthermore, we numerically prove that our DIQRNG protocols are robust to noise. That is, close-to-perfect experiments certify close to $2 \log(d)$ bits of global randomness.

\prlparagraph{Preliminaries}
In a Bell scenario, two parties---Alice and Bob---share a pair of entangled quantum systems and perform measurements on them. Alice chooses a measurement setting $x$ from a finite alphabet and observes an outcome $a$ also from a finite alphabet. Similarly, Bob chooses $y$ and observes $b$. According to quantum theory, the probability of outcomes $a$ and $b$ given settings $x$ and $y$ is given by
\begin{equation}
    p(a,b|x,y) = \tr[ (A^x_a \otimes B^y_b) \rho ],
\end{equation}
where $\rho$ is a quantum state (positive semidefinite operator with unit trace) on a tensor product Hilbert space $\cH_A \otimes \cH_B$, $(A^x_a)_a$ is a positive operator-valued measure (POVM) on $\cH_A$ for every $x$ (that is, $A^x_a$ are positive semidefinite operators adding up to the identity on $\cH_A$) and $(B^y_b)_b$ is a POVM on $\cH_B$ for every $y$. In this work, we assume Hilbert spaces to be finite-dimensional.

The aim of DIQRNG is to certify randomness based only on the observed probabilities (the correlation). In particular, the amount of randomness certifiable from the outcomes of the measurements $x$ and $y$ can be quantified using the classical-quantum state \cite{TCR09}
\begin{equation}\label{eq:ccq_state}
\begin{split}
    \sigma^{x,y}_{\rA \rB E} = & \left. \sum_{a,b} \ketbraq{a}_\rA \otimes \ketbraq{b}_\rB \right. \\
    & \left. \otimes \tr_{AB}[ ( A^x_a \otimes B^y_b \otimes \I_E ) \ketbraq{\psi}],
    \right.
\end{split}
\end{equation}
where $\{ \ket{a}_\rA \}_a$ is an orthonormal basis of a Hilbert space with the same dimension as the number of possible outcomes of Alice, and $\{ \ket{b}_\rB \}_b$ is similarly defined (note that these Hilbert spaces are denoted using an upright $\rA$ and $\rB$, whereas the Hilbert spaces of the quantum state are italic $A$ and $B$). Furthermore, $\ket{\psi}$ is a purification of the state used to obtain $p(a,b|x,y)$; in other words, we have $p(a,b|x,y) = \bra{\psi}(A^x_a \otimes B^y_b \otimes \I_E) \ket{\psi}$.

Crucially, in DIQRNG we don't have information about the underlying state ($\rho$ or $\ket{\psi}$) and measurements ($A^x_a$ and $B^y_b$). All claims are made based only on the observed correlation. The device-independent randomness is thus characterised by the conditional von-Neumann entropy $H(\rA,\rB|E)$ of the state in Eq.~\eqref{eq:ccq_state}, minimised over all states $\ket{\psi}$ and all POVMs $A^x_a$ and $B^y_b$ compatible with the observed correlation \cite{TCR09}. The intuition behind this formula is that it quantifies the uncertainty in the measurement outcomes of Alice and Bob, given quantum side information of a potential eavesdropper Eve, who holds a purification of the state shared between Alice and Bob. This uncertainty then has to be minimised over all possible realisations of the observed correlation in order to obtain the most conservative randomness estimate. When the correlation arises from a locally $d$-dimensional state (i.e.~$\dim \cH_A = \dim \cH_B = d$) and projective measurements, then $H(\rA,\rB|E)$ cannot be larger than $2 \log(d)$, since projective measurements have at most $d$ outcomes.

To bound device-independent randomness, we often use a \emph{Bell inequality}~\cite{brunner2014bell}, which is a linear functional of correlations, $\hat{W}(p) = \sum_{a,b,x,y} c_{abxy} p(a,b|x,y)$ for some $c_{abxy} \in \mathbb{R}$. The value of a Bell inequality for a given state $\rho$ and collection of POVMs $\cA = ((A^x_a)_a)_x$ and $\cB = ((B^y_b)_b)_y$ can be expressed in terms of the \emph{Bell operator}, $W(\cA, \cB) = \sum_{a,b,x,y} c_{abxy} A^x_a \otimes B^y_b$, through $\hat{W}(p) = \tr[W(\cA, \cB) \rho]$. In certain cases, observing the maximal possible value of a Bell inequality certifies useful information about the state and measurements used in the experiment. An often used technique to obtain such certification is a \emph{sum-of-squares} (SOS) decomposition. An SOS decomposition~\cite{bamps2015sum,vsupic2016self} is an operator inequality satisfied by the Bell operator which has the form
\begin{equation}\label{eq:SOS}
    \beta \I - W(\cA, \cB) \ge \sum_j P_j^\dagger(\cA,\cB) P_j(\cA,\cB)
\end{equation}
for some $\beta \in \mathbb{R}$ and operator-valued functions $P_j$, where the operator inequality holds under the assumption that $\cA$ and $\cB$ are collections of POVMs. Note that the right-hand side of Eq.~\eqref{eq:SOS} is positive semidefinite by design, and therefore constructing an SOS decomposition proves that $\hat{W}(p) = \tr[W(\cA,\cB) \rho] \le \beta$ for any state $\rho$ and measurements $\cA$ and $\cB$.

Moreover, if $\hat{W}(p) = \beta$ can be attained with some state $\rho$ and POVMs $\cA$ and $\cB$, then these must satisfy $\tr[ P_j^\dagger(\cA,\cB) P_j(\cA,\cB) \rho] = 0$ which also implies $P_j(\cA,\cB) \rho = 0$ for all $j$. These equations constrain the state and measurements that give rise to the maximal value of the Bell inequality. In some cases, these constraints are sufficient to compute the conditional entropy of all classical-quantum states of the form in Eq.~\eqref{eq:ccq_state} that are compatible with the observed correlations.

\prlparagraph{Results}
To certify $2 \log(d)$ bits of global randomness in every dimension $d$ using locally $d$-dimensional states and projective measurements, we use a variant of a family of Bell inequalities originally introduced in Ref.~\cite{TFRB+21} . The Bell inequalities are parametrised by an integer $d \ge 2$. In the ideal realisation, $d$ is also the local dimension of the state, but importantly, $d$ is simply a parameter of the Bell inequality. In the Bell scenario, Alice has two settings, $x \in \{1,2\}$ with $d$ outcomes each labelled by $a \in \{1, 2, \ldots, d\} =: [d]$. Bob has $d^2+1$ settings. The first $d^2$ are labelled by $jk$ where $j,k \in [d]$, and each of these have three outcomes, labelled by $b \in \{1,2,3\}$. Bob's last measurement setting is labelled by $\rr$ (for ``randomness'') and this has $d$ outcomes. Alice and Bob aim to generate randomness from their settings `2' and `$\rr$', respectively.

The Bell inequality used to certify randomness is
\begin{equation}\label{eq:Bell}
\begin{split}
\hat{W}_d(p) = & \left. \sum_{j,k=1}^{d} \Big[ p(j,1|1,jk) - p(j,2|1,jk) \right. \\
& \left. + p(k,2|2,jk) - p(k,1|2,jk) \Big] \right. \\
& \left. - \frac12 \sqrt{ \frac{d-1}{d} } \sum_{j,k=1}^{d} \big[ p_B(1|jk) + p_B(2|jk) \big] \right. \\
& \left. + \sum_{j=1}^{d} p(j,j|1,\rr),
\right.
\end{split}
\end{equation}
where $p_B$ is Bob's marginal distribution.
The first two sums correspond exactly to the Bell inequalities from Ref.~\cite{TFRB+21}. The new element is the addition of the setting $\rr$ to Bob's side, which features in the last term of $\hat{W}_d$.

From Ref.~\cite{TFRB+21}, we know that the maximal quantum value achievable for the first two sums is $\sqrt{d(d-1)}$ and this can be reached by a locally $d$-dimensional state and projective measurements. This implies that the maximal quantum value of $\hat{W}_d$ is $\sqrt{d(d-1)} + 1$. In order to reach this maximal value, the first two sums must be maximised, which certifies the following properties \cite{TFRB+21,PAK22,Far24}: for every purification $\ket{\psi}_{ABE}$ of the bipartite state $\rho$ on $\cH_A \otimes \cH_B$ used to reach the maximal violation, there exist Hilbert spaces $\cH_{A'}$ and $\cH_{B'}$, local isometries $V_A : \cH_A \to \bC^d \otimes \cH_{A'}$ and $V_B : \cH_B \to \bC^d \otimes \cH_{B'}$ such that
\begin{align}\label{eq:selftest1}
(V_A \otimes V_B \otimes \I_E)(A^1_j \otimes \I_B \otimes \I_E)\ket{\psi}_{ABE} \\ 
\nonumber
= (\ketbraq{j} \otimes \I_{\bC^d}) \ket{\phi^+_d} \otimes \ket{ \aux }_{A'B'E},
\end{align}
where $\{ \ket{j} \}_{j=1}^d$ is the computational basis on $\bC^d$ and $\ket{\phi^+_d} = \frac{1}{\sqrt{d}} \sum_{j=1}^d \ket{j} \otimes \ket{j} \in \bC^d \otimes \bC^d$ is the maximally entangled state. Eq.~\eqref{eq:selftest1} also implies (summing over $j$) that
\begin{equation}\label{eq:selftest2}
(V_A \otimes V_B \otimes \I_E)\ket{\psi}_{ABE} = \ket{\phi^+_d} \otimes \ket{ \aux }_{A'B'E}.
\end{equation}
Furthermore, Alice's measurements satisfy the algebraic relations (when acting on the state)
\begin{align}\label{eq:selftest3}
(A^1_j A^2_k A^1_j \otimes \I_B \otimes \I_E) \ket{\psi}_{ABE} = \frac1d (A^1_j \otimes \I_B \otimes \I_E) \ket{\psi}_{ABE}
\end{align}
for all $j,k \in [d]$. Summing the above over $k$, we also see that $A^1_j$ act on the state as projections, that is,
\begin{align}\label{eq:selftest4}
( ( A^1_j )^2 \otimes \I_B \otimes \I_E) \ket{\psi}_{ABE} = (A^1_j \otimes \I_B \otimes \I_E) \ket{\psi}_{ABE}.
\end{align}
In particular, any pair of $d$-dimensional mutually unbiased bases \cite{MUBreview} satisfies these relations.

The last term in $\hat{W}_d$ in Eq.~\eqref{eq:Bell} is bounded by 1, and the Bell operator corresponding to this term is $\sum_{j=1}^d A^1_j \otimes B^\rr_j$. Notice that
\begin{equation}
(A^1_j \otimes \I - \I \otimes B^\rr_j)^2 = (A^1_j)^2 \otimes \I + \I \otimes (B^\rr_j)^2 - 2A^1_j \otimes B^\rr_j,
\end{equation}
and therefore
\begin{equation}
A^1_j \otimes B^\rr_j = \frac12\big[ (A^1_j)^2 \otimes \I + \I \otimes (B^\rr_j)^2 \big] - \frac12 (A^1_j \otimes \I - \I \otimes B^\rr_j)^2.
\end{equation}
This leads to the operator inequality
\begin{equation}\label{eq:SOS_lastbit}
\begin{split}
    & \left. \sum_{j=1}^d A^1_j \otimes B^\rr_j = \frac12 \sum_{j=1}^{d} \big[ (A^1_j)^2 \otimes \I + \I \otimes (B^\rr_j)^2 \big] \right. \\
    & \left. \quad - \sum_{j=1}^{d}\frac12 (A^1_j \otimes \I - \I \otimes B^\rr_j)^2 \right. \\
    & \left. \le \frac12 \sum_{j=1}^{d} \big[ A^1_j \otimes \I + \I \otimes B^\rr_j \big] - \sum_{j=1}^{d}\frac12 (A^1_j \otimes \I - \I \otimes B^\rr_j)^2 \right. \\
    & \left. = \I - \sum_{j=1}^{d}\frac12 (A^1_j \otimes \I - \I \otimes B^\rr_j)^2
    \right.
\end{split}
\end{equation}
where the inequality comes from the fact that $(A^1_j)^2 \le A^1_j$ and $(B^\rr_j)^2 \le B^\rr_j$. Rearranging Eq.~\eqref{eq:SOS_lastbit} is an SOS decomposition for the last term in the Bell inequality, $\I - \sum_{j=1}^d A^1_j \otimes B^\rr_j \ge  \sum_{j=1}^{d}\frac12 (A^1_j \otimes \I - \I \otimes B^\rr_j)^2$. To reach the maximal value, we therefore need that for all $j \in [d]$ we have  $(A^1_j \otimes \I)\rho = (\I \otimes B^\rr_j) \rho$. Since in finite dimensions all purifications are locally isometric on the purifying Hilbert space, this implies that for any purification $\ket{\psi}_{ABE}$ of $\rho$, we have
\begin{equation}\label{eq:selftest5}
(A^1_j \otimes \I \otimes \I) \ket{\psi}_{ABE} = (\I \otimes B^\rr_j \otimes \I) \ket{\psi}_{ABE} \quad \forall j \in [d].
\end{equation}
Note that this together with Eq.~\eqref{eq:selftest4} implies that $B^\rr_j$ are projective on the state as well, i.e.~
\begin{equation}\label{eq:selftest6}
(\I \otimes (B^\rr_j)^2 \otimes \I) \ket{\psi}_{ABE} = (\I \otimes B^\rr_j \otimes \I) \ket{\psi}_{ABE} \quad \forall j \in [d].
\end{equation}

The maximal violation of the Bell inequality in Eq.~\eqref{eq:Bell} therefore leads to the certification results in Eqs.~\eqref{eq:selftest1}--\eqref{eq:selftest4}, and \eqref{eq:selftest5}--\eqref{eq:selftest6}. These turn out to be sufficient to bound the device-independent randomness. In particular, for any collection of POVMs $\cA$, $\cB$ and any state with purification $\ket{\psi}_{ABE}$ that are compatible with the maximal violation, the classical-quantum state \eqref{eq:ccq_state} corresponding to the measurements `2' for Alice and `$\rr$' for Bob has the form
\begin{widetext}
\begin{equation}
\begin{split}
\sigma^{2, \rr}_{\rA \rB E} & \left. = \sum_{a,b=1}^{d} \ketbraq{a}_\rA \otimes \ketbraq{b}_\rB \otimes \tr_{AB} [ (A^2_a \otimes B^\rr_b \otimes \I_E) \ketbraq{\psi}_{ABE} ] \right. \\
& \left. = \sum_{a,b=1}^{d} \ketbraq{a}_\rA \otimes \ketbraq{b}_\rB \otimes \tr_{AB} [ (A^2_a \otimes B^\rr_b \otimes \I_E) \ketbraq{\psi}_{ABE} (\I_A \otimes B^\rr_b \otimes \I_E)] \right. \\
& \left. = \sum_{a,b=1}^{d} \ketbraq{a}_\rA \otimes \ketbraq{b}_\rB \otimes \tr_{AB} [ (A^2_a A^1_b \otimes \I_B \otimes \I_E) \ketbraq{\psi}_{ABE} (A^1_b \otimes \I_B \otimes \I_E)]  \right. \\
& \left. = \sum_{a,b=1}^{d} \ketbraq{a}_\rA \otimes \ketbraq{b}_\rB \otimes \tr_{AB} [ (\frac1d A^1_b \otimes \I_B \otimes \I_E) \ketbraq{\psi}_{ABE}]
\right.
\end{split}
\end{equation}
\end{widetext}
Here, the second equation is a consequence of the cyclicty of the partial trace and the projectivity of $B^\rr_b$ when acting on $\ket{\psi}_{ABE}$. The third equation is Eq.~\eqref{eq:selftest5} applied twice, and the fourth one is Eq.~\eqref{eq:selftest3} together with the cyclicity of the partial trace. We now plug in the isometries $V_A^\dagger V_A \otimes V_B^\dagger V_B = \I_A \otimes \I_B $ from Eq.~\eqref{eq:selftest1} to obtain
\begin{widetext}
\begin{equation}
\begin{split}
\sigma^{2, \rr}_{\rA \rB E} & \left. = \frac1d \sum_{a,b=1}^{d} \ketbraq{a}_\rA \otimes \ketbraq{b}_\rB \otimes \tr_{AB} [ (V_A^\dagger V_A A^1_b \otimes V_B^\dagger V_B \otimes \I_E) \ketbraq{\psi}_{ABE}] \right. \\
& \left. = \frac1d \sum_{a,b=1}^{d} \ketbraq{a}_\rA \otimes \ketbraq{b}_\rB \otimes \tr_{\bC^d \otimes \bC^d \otimes A' B'} [ ( V_A A^1_b \otimes V_B \otimes \I_E) \ketbraq{\psi}_{ABE} (V_A^\dagger \otimes V_B^\dagger \otimes \I_E)] \right. \\
& \left. = \frac1d \sum_{a,b=1}^{d} \ketbraq{a}_\rA \otimes \ketbraq{b}_\rB \otimes \tr_{\bC^d \otimes \bC^d \otimes A' B'} [ (\ketbraq{b} \otimes \I_{\bC^d} ) \ketbraq{\phi^+_d} \otimes \ketbraq{ \aux }_{A'B'E} ] \right. \\
& \left. = \frac{1}{d^2} \sum_{a,b=1}^{d} \ketbraq{a}_\rA \otimes \ketbraq{b}_\rB \otimes \tr_{A'B'} ( \ketbraq{ \aux }_{A'B'E} ) = \frac{\I_{\bC^{d^2}}}{d^2} \otimes \eta_E.
\right.
\end{split}
\end{equation}
\end{widetext}
Here, the second equality is the cyclicity of the partial trace, the third one is Eqs.~\eqref{eq:selftest1} and \eqref{eq:selftest2}, the fourth one is taking the partial trace on $\bC^d \otimes \bC^d$, and in the last we introduce $\eta_E := \tr_{A'B'} ( \ketbraq{ \aux }_{A'B'E} )$. It is straightforward to see that for any such classical-quantum state $\sigma^{2, \rr}_{\rA \rB E}$, we have $H(\rA,\rB|E) = \log(d^2) = 2 \log(d)$. In other words, if Alice and Bob observe the maximal violation of the Bell inequality $\hat{W}_d$, then they can certify the theoretical maximum in dimension $d$, $2\log(d)$ bits of randomness from projective measurements. Importantly, this maximum can be achieved by systems of local dimension $d$, as shown in Ref.~\cite{TFRB+21}, supplemented by $B^\rr_j = (A^1_j)^T$, where $(.)^T$ is transposition in the computational basis.

\prlparagraph{Robustness}
Observing the maximal violation exactly is practically infeasible. To gauge the practicality of our protocol, we numerically compute lower bounds on the device-independent randomness in realistic cases when the violation is not exactly maximal. To obtain these bounds, we first lower bound the von Neumann entropy $H(\rA,\rB|E)$ with the min-entropy, and then compute bounds on the min-entropy~\cite{konig2009operational} using established semidefinite programming techniques~\cite{10.1088/978-0-7503-3343-6,mironowicz2024semi,tavakoli2024semidefinite} with the Navascu{\'e}-Pironio-Ac{\'\i}n (NPA) method~\cite{navascues2007bounding,navascues2008convergent,navascues2015almost}.

\begin{figure}[t!]
    \centering
    \includegraphics[width=0.48\textwidth]{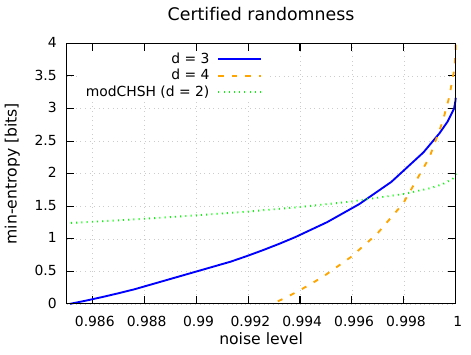}
    \caption{
        Certified randomness (min-entropy) as a function of the noise level $\eta$ defined in~\eqref{eq:noise_level}.
        The curves correspond to $d=3$ (blue, solid), $d=4$ (orange, dashed), and the modCHSH inequality (green, dotted).
        For $d=3$ and $d=4$, certified randomness exceeds 2 bits around $\eta \approx 0.997$ and $\eta \approx 0.998$, respectively, demonstrating an advantage over binary-outcome ($d=2$) scenarios.
        The plot was obtained using the Navascués--Pironio--Acín (NPA) hierarchy at the level 1+AB.
    }
    \label{fig:minentropy_vs_violation}
\end{figure}

Figure~\ref{fig:minentropy_vs_violation} shows the resulting min-entropy as a function of the noise level for the three considered Bell certificates. Two of these are from the family discussed in this work, given by Eq.~\eqref{eq:Bell} for $d=3$ and $d=4$, respectively. The so-called modCHSH inequality is included for comparison. It corresponds to the Bell inequality (and corresponding DIQRNG protocol) introduced in~\cite{mironowicz2013robustness}, where Alice has three measurement settings and Bob has two, each with binary outcomes. The maximal violation of this Bell inequality can be achieved with local dimension $d=2$. We included this Bell inequality as it is known for its high robustness among binary outcome inequalities.

On the horizontal axis we plot the \emph{noise level}, defined as a normalised measure of the violation of a Bell expression. 
For each Bell certificate, the normalisation is performed using its maximal quantum value ($\beta_Q$) and its white-noise value ($\beta_{\mathrm{WN}}$, obtained by assigning equal probability to all outcomes).
For the Bell inequalities proposed in this work, $\beta_Q=\sqrt{d(d-1)} + 1$, while for the modCHSH $\beta_Q=1 + 2\sqrt{2}$. 
For the Bell inequalities introduced here, the white-noise values are $\beta_{\mathrm{WN}}(d=3) \approx -2.1162$, and $\beta_{\mathrm{WN}}(d=4) \approx -4.3688$.
Using these quantities, the noise level $\eta$ is defined as
\begin{equation}
	\label{eq:noise_level}
    \eta \equiv \frac{\beta_{\mathrm{obs}} - \beta_{\mathrm{WN}}}{\beta_Q - \beta_{\mathrm{WN}}},
\end{equation}
where $\beta_{\mathrm{obs}}$ denotes the observed value of the Bell expression.
The observed Bell value $\beta_{\mathrm{obs}}$ is imposed as a constraint in the NPA optimisation at level $1+AB$. For each value of the noise level, we upper-bound the corresponding guessing probability $P_{\mathrm{guess}}$ of the outcomes used to generate randomness. The certified randomness is quantified by the min-entropy, $H_{\min} = -\log_2 P_{\mathrm{guess}}$.

In the case of maximal violation, all three cases ($d=2,3,4$) yield the maximum possible global randomness of $2 \log (d)$ bits, consistent with the theoretical limit. In the near-threshold regime ($\eta \approx 0.993$), we observe a gain in certified randomness for dimensions $d>2$ compared to the modCHSH case. In particular, for $d=3$ and $d=4$, the curves surpass 2 bits of certified randomness at $\eta \approx 0.997$ and $\eta \approx 0.998$, respectively, thus surpassing the maximal amount of certifiable global randomness for $d=2$.
Remarkably, experimental noise levels given by Eq.~\eqref{eq:noise_level} of the order of $0.9975$ have been reported for qubit systems~\cite{seguinard2023experimental}, suggesting that the advantages of using higher-dimensional systems ($d>2$) for device-independent randomness certification are within reach of current technology.

\prlparagraph{Conclusions and outlook}
We have shown that for every dimension $d$, the theoretical maximum, $2\log(d)$ bits of global device-independent randomness can be certified from a system of local dimension $d$ using projective measurements. We introduced explicit protocols reaching this bound, and numerically prove the protocols' robustness to experimental imperfections.

To make full use of these protocols, analytic techniques for robust certification (such as robust self-testing) would be desirable. It further remains open how much randomness can be certified from a system of local dimension $d$ if only one party is restricted to projective measurements. To tackle this question, one would need to devise methods for finding a non-projective measurement that maximises device-independent randomness for a fixed projective measurement for the other party.

\prlparagraph{Acknowledgements}
NPA optimisation was implemented using the Python library ncpol2sdpa~\cite{Wittek_2015} and the MOSEK solver~\cite{mosek}. We acknowledge the use of a computational server financed by the Foundation for Polish Science (IRAP project, ICTQT, contract no. 2018/MAB/5/AS-1, co-financed by EU within Smart Growth Operational Programme). The Center for QuantumEnabled Computing project is carried out within the
International Research Agendas programme of the Foundation for Polish Science co-financed by the European
Union under the European Funds for Smart Economy
2021-2027 (FENG).

\prlparagraph{Note added} Notice that similar results were independently developed in Ref.~\cite{Barcelona} where novel constructions of Bell inequalities for global randomness certification were provided.


\bibliography{bib_2logd_projective}

%
%
%
%

\end{document}